# Quantum tomography based on principles of completeness, adequacy and fidelity

Yu. I. Bogdanov[a,b,c], N. A. Bogdanova[a,b], B. I. Bantysh[a,b], D. V. Fastovets[a,b], V. F. Lukichev[a]

[a]Valiev Institute of Physics and Technology, Russian Academy of Sciences, Russian Federation;
[b]National Research University of Electronic Technology (MIET), Russian Federation;
[c]National Research Nuclear University (MEPhI), Russian Federation

## ABSTRACT

In this report we present a general approach for estimating quantum circuits by means of measurements. We apply the developed general approach for estimating the quality of superconducting and optical quantum chips. Using the methods of quantum states and processes tomography developed in our previous works, we have defined the adequate models of the states and processes under consideration.

## 1. INTRODUCTION

At present dozens of various models of quantum computers are being actively discussed. Among the most prospective and interesting suggestions are the projects based on superconducting structures, photons, atom and ion traps and other [1-8]. The main achievement of the research in the field performed until now has been a practical demonstration of validity of physical principles underlying the idea of quantum computations.

Recently, IBM company started to provide an open access to some of its superconducting quantum processors[9]. However, the quality of these processors is still far from perfect. In present work we implemented the quantum tomography of some operations of IBM quantum processor. Using the methods of quantum states and processes tomography developed in our previous works, we have defined the adequate models of the states and processes under consideration. Some additional important results of this study are presented in our other article in this Proceedings, entitled "High-fidelity quantum tomography with imperfect measurements".

An important motivation of our present work is the development of the most adequate, complete and accurate methods for estimating optical quantum chips by means of quantum measurements. At present, optical quantum chips are being developed in a number of laboratories around the world[10-14]. Active work in this direction is conducted by the team of Professor Kulik in the Center of Quantum Technologies of Moscow State University.

According to the ancient legend, the Earth is located on the backs of three whales. Our approach to tomography of quantum states and quantum operations is also based on three principles (the three whales): completeness, adequacy and fidelity[15,16] (Fig.1).

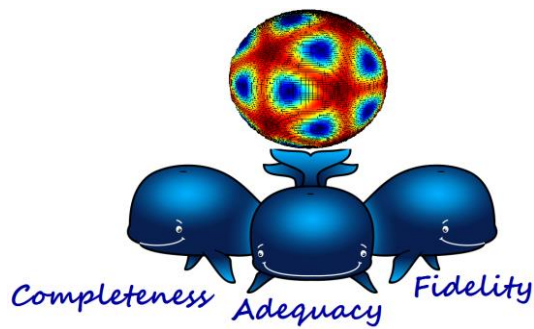

Figure 1. The three whales of quantum tomography.

Completeness means that the quantum measurement protocol allows us to reconstruct any quantum state and any density matrix. To test the completeness property, we introduced a special matrix, based on the operators of quantum measurements[15,16,17]. Completeness is valid if and only if the considered measurement matrix is a matrix of full rank. Adequacy means consistency between the obtained experimental data and the developed quantum model. Usually, we conduct a verification of adequacy through a chi-square test[18,19]. Precision of quantum tomography can be defined by a parameter called fidelity. Fidelity shows how close the reconstructed state is to the ideal theoretical state. Our method based on purification procedure allows us to formulate a generalized statistical distribution for fidelity[18]. This distribution is a natural generalization of the chi-squared distribution.

For the analysis of incomplete measurement protocols we introduced the concept of adjusted fidelity. The concept of adjusted fidelity gives us a way to estimate only those parameters of state, information about which is contained in the experimental data. In this case, we ignore the parameters, information about which is not contained in the measurement protocol at all. To find adjusted fidelity, we use the matrix of complete information for the purified state introduced in our works[18,20].

The plan of our report is as follows.

In the section 2 we consider different ways of describing quantum operations, including operator-sum, unitary representation, and the Choi-Jamiolkowski states with corresponding chi-matrixes.

Then we consider the tomography on the IBM quantum processor. By the example of the analysis of quantum operations, we illustrate three basic principles of tomography: completeness, adequacy, and fidelity.

The next section is devoted to the analysis of linear-optical quantum chips. We consider various protocols of quantum tomography, including protocols that do not satisfy the criterion of completeness.

The final section summarizes the main conclusions of our work.

## 2. QUANTUM OPERATIONS

It is well known that an ideal quantum logical element (gate) performs a unitary transformation of a quantum state (density matrix):

$$\rho_{out} = U \rho_{in} U^\dagger. \quad (1)$$

However, real evolution is never purely unitary. For more realistic models we need to consider the unavoidable interaction of the quantum system with its environment (quantum noise). The relation between the input and the output states is defined by the formula:

$$\rho_{out} = \sum_k E_k \rho_{in} E_k^\dagger \quad (2)$$

where $E_k$ are the so-called Kraus operators.

The operators $E_k$ in the $s$-dimensional Hilbert space can be represented by $s \times s$-matrices. For a unitary transformation there is only one summand defined by the operator $U$ in the sum. Operators $E_k$ must satisfy the constraint to preserve the density matrix trace:

$$\sum_k E_k^\dagger E_k = I, \quad (3)$$

where $I$ is the identity matrix of the order $s$.

Non-unitary transformation (2) acting in a Hilbert space of dimension $s$ can be interpreted as a consequence of some unitary transform $U$ in a higher dimensional space. Let us consider $m$ operators $E_k$ ($k=1,...,m$). Then the $ms \times ms$ dimensional unitary matrix $U$ can be written in the following block form:

$$U = \begin{pmatrix} E_1 & : & : \\ : & : & : \\ E_m & : & : \end{pmatrix}. \tag{4}$$

Here the operators $E_k$ define only the first block-column of the matrix. We can complement the matrix to a unitary by orthogonal complement.

To describe the reduced dynamics of open quantum systems, the concept of complete positivity is used. This concept was proposed and studied by A. S. Holevo in the quantum theory of communication[21], and also by other researchers in statistical mechanics[22-25].

We may recast the property above in another important form based on the use of an ancillary state (ancilla) and Choi-Jamiolkowski state. Let the considered quantum operation $\mathbf{E}$ act on the s-dimensional system $A$. Let us add an ancillary s-dimensional system $B$ and consider the joint system $AB$. Then let us input the maximally entangled state

$$|\Phi\rangle = \frac{1}{\sqrt{s}} \sum_{j=1}^{s} |j\rangle \otimes |j\rangle \tag{5}$$

to our operation. Here the first factor in tensor product corresponds to the subsystem A, and the second corresponds to the subsystem B.

The output state of such process is called the Choi-Jamiolkowski state. Let the identical transform $\mathbf{I}$ act on the subsystem B. Then the transform $(\mathbf{I} \otimes \mathbf{E})$ acts on the entire system $AB$.

It appears that if the density matrix $(|\Phi\rangle\langle\Phi|)$ is submitted to the input, then the chi-matrix is obtained at the output. However, the trace of this matrix is equal to 1, i.e.

$$(\mathbf{I} \otimes \mathbf{E})(|\Phi\rangle\langle\Phi|) = \rho_\chi, \text{ where } \rho_\chi = \frac{1}{s}\chi. \tag{6}$$

We illustrate the above considerations by the following figure.

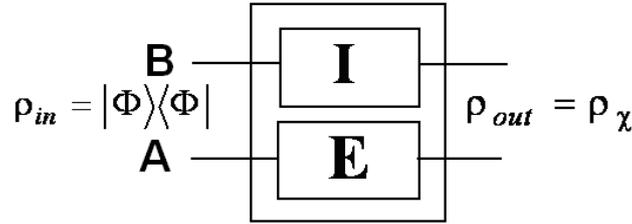

Figure 2. Quantum scheme of Choi-Jamiolkowski state calculation.

## 3. FIDELITY

Precision of quantum tomography can be defined by a parameter called Fidelity[26,27]

$$F = \left(\text{Tr}\sqrt{\rho_0^{1/2} \rho \rho_0^{1/2}}\right)^2 \tag{7}$$

where $\rho_0$ is theoretical density matrix and $\rho$ is reconstructed density matrix.

Fidelity shows how close the reconstructed state is to the ideal theoretical state. The reconstruction is precise if Fidelity is equal to one.

This equation looks quite complex, but it becomes simple if we apply the Uhlmann theorem[26]. According to the theorem, Fidelity is simply the maximum possible squared absolute value of the scalar product, which we may obtain using purification procedure.

$$F = |\langle c_0 | c \rangle|^2 \qquad (8)$$

where $c_0$ and $c$ are theoretical and reconstructed purified state vectors.

We explicitly use the Uhlmann theorem in our algorithm of statistical reconstruction of quantum states. This fact is very important. Even if the state is not pure, we have to purify it by moving into the space of higher dimension[18].

It is well known that purified state vectors are defined ambiguously. However, this ambiguity does not preclude us from reconstructing a quantum state. This is a very useful feature of our algorithm. It is devised in the way that different purified state vectors produce the same density matrix and therefore the same fidelity during the reconstruction. This principle is very important for proposed procedure and thus reconstruction can be held by means of purification. Purification greatly facilitates the search of solution, especially when we need to estimate a great number of parameters (hundreds or even thousands).

It is equally important that due to the usage of purification procedure we succeed in formulating a generalized statistical distribution for fidelity[18]. The value $1-F$ can be called the loss of fidelity. It is a random value and its asymptotical distribution can be presented in the following form:

$$1 - F = \sum_{j=1}^{j_{\max}} d_j \xi_j^2 \qquad (9)$$

where $d_j \geq 0$ are non-negative coefficients, $\xi_j \sim N(0,1)$, $j = 1, \ldots, j_{max}$ are independent normally distributed random values with zero mean and variance equal to one, $j_{\max} = (2s - r)r - 1$ is the number of degrees of freedom of a quantum state and corresponding distribution; $s$ is the Hilbert space dimension, $r$ is the rank of mixed state, which is the number of non-zero eigenvalues of the density matrix. In particular $j_{\max} = 2s - 2$ for pure states and $j_{max} = s^2 - 1$ for mixed states of full rank ($r = s$).

This distribution is a natural generalization of chi-squared distribution. Ordinary Chi-squared distribution corresponds to the particular case when $d_1 = d_2 = \ldots = d_{j_{\max}} = 1$ (all components of vector $d$ are equal to one).

From (3) we get that average fidelity loss is equal to this expression.

$$\langle 1 - F \rangle = \sum_{j=1}^{j_{\max}} d_j \qquad (10)$$

It is also easy to show that the variance for fidelity loss is given by the following equation:

$$\sigma^2 = 2 \sum_{j=1}^{j_{\max}} d_j^2 \qquad (11)$$

In the asymptotical limit considered by us, parameters $d_j$ are inversely proportionate to the sample size $n$. Let us introduce the value of fidelity loss which is independent from sample size.

$$L = n\langle 1-F \rangle = n\sum_{j=1}^{j_{max}} d_j \qquad (12)$$

This quantity will be the main characteristic of precision in examples mentioned later. However, prior to examples it is better to study what the protocol of measurement is.

## 4. QUANTUM TOMOGRAPHY ON IBM QUANTUM PROCESSOR: COMPLETENESS, ADEQUACY AND FIDELITY

This section describes our research on the tomography of the IBM quantum processor. The quantum measurement protocol included 36 rows: 6 projection states at the input (eigenvectors of Pauli matrices) and the same 6 projection states at the output. The total sample size was more than 18 thousand representatives.

Completeness means that the quantum measurement protocol allows us to reconstruct any quantum state and any density matrix. To test the completeness property, we introduced a special matrix $B$, based on the operators of quantum measurements[15,16,17]. Completeness is valid if and only if the considered measurement matrix is a matrix of full rank (all singular values are strictly positive).

Verification of the completeness of the considered quantum protocol:

$$svd(B) = [3\ \sqrt{3}\ \sqrt{3}\ \sqrt{3}\ \sqrt{3}\ \sqrt{3}\ \sqrt{3}\ 1\ 1\ 1\ 1\ 1\ 1\ 1\ 1\ 1], \qquad rank(B) = 16.$$

Adequacy means consistency between the obtained experimental data and the developed quantum model. Usually we conduct a verification of adequacy through a chi-square test.

In our case: $\chi^2 = 18.002$, number of degrees of freedom = 10, p-value = 5.5%

Fidelity with respect to ideal transform $F(|\psi_{ideal}\rangle, \rho_{true}) = \langle \psi_{ideal} | \rho_{true} | \psi_{ideal} \rangle$. In our case: $F \approx 93,29\%$.

Some additional results of reconstruction of the gate Z are shown on the Fig. 3. The results include the estimation of an adequate rank of the quantum operation, the statistical reconstruction of the chi-matrix, and the estimation of the Fidelity with respect to ideal transform.

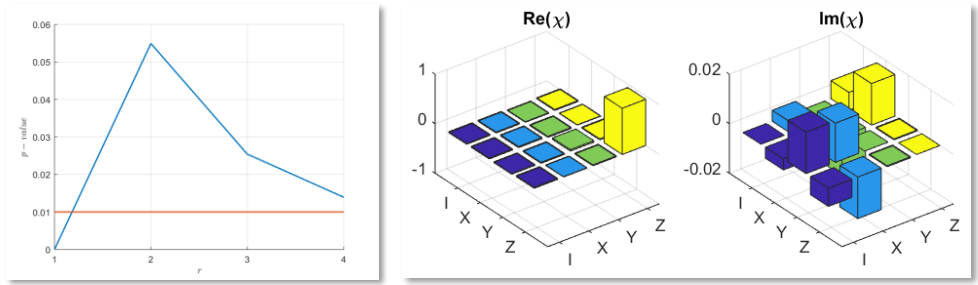

Figure 3. Estimation of an adequate rank (left). The reconstructed chi-matrix in the representation of the Pauli operators, rank 2 (right).

The following Fig.4 shows the fiducial distribution of fidelity. This distribution answers the following question: how close is the reconstructed quantum operation to an unknown quantum operation. From the figure, we see that the agreement is really very high. With a confidence of 95%, it can be argued that the fidelity of the model with respect to the unknown exact operation is at least 0.9984.

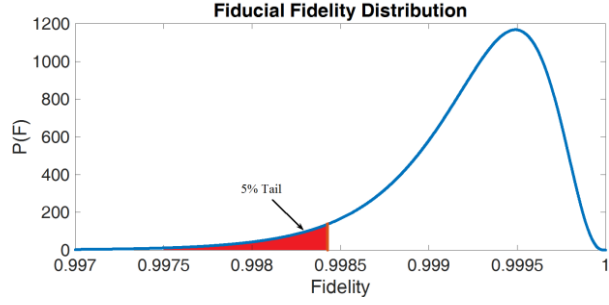

Figure 4. Z gate tomography: Fiducial Fidelity Distribution. Fidelity of the model with respect to the unknown exact operation > 0.9984 with confidence of 95%.

## 5. OPTICAL QUANTUM CHIPS MEASURING

This section describes our approach to measuring optical quantum chips. We will consider two different types of measurements[11]. The first type of measurement reduces to an estimate of transition probabilities. We consider the probability that a photon from some given mode at the input will be in some given mode at the output. The second type of measurement is based on states in which a photon is represented in a superposition of two modes with a given phase difference. Such states are prepared by using a beam-splitter and a phase shifter to control the relative phase between states.

We consider three different protocols of quantum measurements.

The simplest protocol is reduced to measuring only transition probabilities.

Set № 1: $|j\rangle$ - photon in the mode $j$, $j = 1,...,N$

It is clear that this protocol does not satisfy the criterion of completeness, which was mentioned above. In order to build a complete protocol, we add projections to the states in which the photon is represented by superpositions of various modes. We will consider superpositions of all possible pairs of modes, both at the input and at the output of the optical quantum network.

Extended protocol of quantum measurements.

Set № 1 + Set № 2: $|j,k\rangle = \frac{1}{\sqrt{2}}\left(\exp(-i\varphi/2)|j\rangle + \exp(i\varphi/2)|k\rangle\right)$, $j = 1,...,N-1$, $k = j+1,...,N$, $0 \leq \varphi < 2\pi$.

Such an extended protocol satisfies the criterion of completeness, which is confirmed by calculations. Implementation of such a protocol, however, requires significant resources.

The third protocol that we consider occupies an intermediate position. This protocol uses only a limited set of projective states.

It is assumed that at the input only the first mode is capable of creating superpositions with other modes. In addition, it is assumed that at the output, superpositions between modes are excluded altogether. In the extended protocol, on the contrary, superpositions are possible between any pair of modes, both at the input and at the output. Note that the third (restricted) protocol also does not satisfy the completeness criterion, as is the simplest protocol (although it contains substantially more information).

Restricted intermediate protocol: at the input in the set №2 we use only $j = 1$; at the output we use only set №1.

The following Figure 5 illustrates the use of the concept of adjusted fidelity, which we introduced for the analysis of incomplete measurement protocols. In the example on the figure, the initial accuracy is very low, less than 0.2. At the same time, the adjusted fidelity is very high, more than 0.9999. The concept of adjusted fidelity gives us a way to estimate only those parameters of state, information about which is contained in the experimental data. In this case, we ignore the parameters, information about which is not contained in the measurement protocol at all. To find adjusted fidelity, we use the matrix of complete information for the purified state introduced in our works. This matrix is given in a real Euclidean space of double dimension. If the protocol of measurements is complete, then all physically significant Euclidean space is measurable.

If the protocol of measurements is incomplete, then there are two subspaces: a measurable subspace and an unmeasurable subspace. To calculate adjusted fidelity, we take into account only the projection onto the measurable subspace of a real Euclidean space of double dimension and ignore the projection on the unmeasurable subspace.

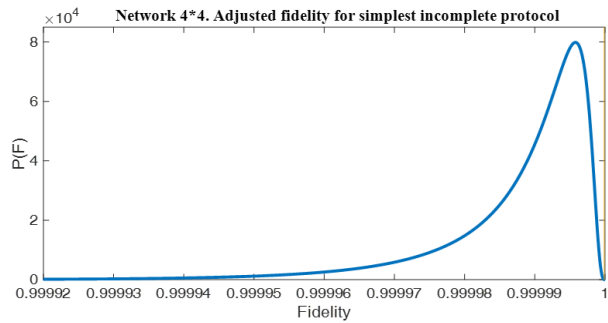

Figure 5. Adjusted fidelity. Network 4*4. Simplest incomplete protocol. Initial fidelity=0.199. Adjusted fidelity>0.9999. Verification of adequacy using the chi-square test: chi_2 = 7.14, number of degrees of freedom = 6, p-value =31%

The concepts of completeness, adequacy and fidelity, as well as the concept of adjusted fidelity, can be applied to arbitrary optical chips. This is illustrated by the following two figures. The figures compare a restricted incomplete protocol and an extended complete protocol. The Fig. 6 shows a tomography of a noisy optical quantum network 4-by-4.

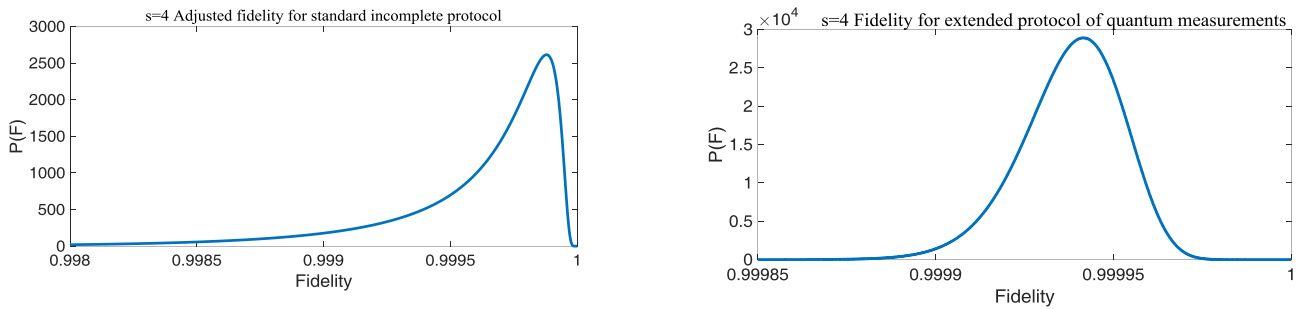

Figure 6. Noisy Quantum Network 4*4: restricted vs extended protocol. Restricted protocol (left) is incomplete: rank(B) = 40. Using fidelity adjusted to incomplete protocol of measurements. Extended protocol (right) is complete: rank(B) = 256. Rank of quantum operation =2

The Fig. 7 examines the tomography of a unitary quantum network 8-by-8. We note that in the case under consideration the measurements matrix of the process has a rather high dimensionality: more than 10000 rows and more than 4000 columns.

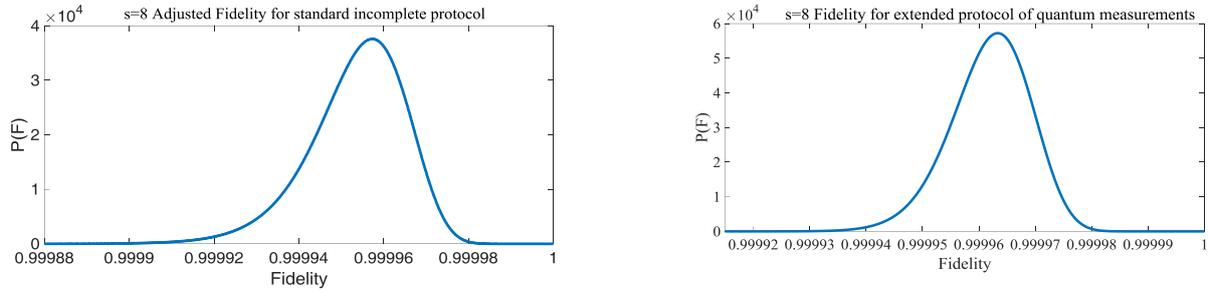

Figure 7. Unitary Quantum Network 8*8: restricted vs extended protocol. Restricted protocol (left) is incomplete: rank(B) =176. Using fidelity adjusted to incomplete protocol of measurements. Extended protocol is complete: rank(B) = 4096

## 6. CONCLUSIONS

In conclusion, we present the main results.

A general approach for estimating quantum schemes by means of measurements is proposed.

A developed approach is applied to estimation the quality of superconducting and optical quantum chips.

The methods and algorithms developed within the framework of this analysis provide an effective means of controlling various quantum operations.

The obtained results are of practical value for the tasks of provision of quality and effectiveness of technologies for building quantum computers and simulators.

## ACKNOWLEDGEMENT

The work was supported by the Russian Science Foundation.

Authors are grateful to Dr. Konstantin Katamadze for his discussion and cooperation